# The J/Ψ meson and the missing heavy baryon octet

Kai-Wai Wong[1], Gisela A. M. Dreschhoff[1], Högne J. N. Jungner[2]

[1]Department of Physics and Astronomy, University of Kansas, Lawrence, Kansas, USA, [2] Radiocarbon Dating Lab, University of Helsinki, Helsinki, Finland

The 5D homogeneous space-time projection theory produces the Gell-Mann standard model, and the gluon fields together with quantum gauge constraint which is responsible for the major portion of the hadron mass as discussed previously. It was found that the SU(3) representations for the mesons and baryons together with the mass levels within each representations are generated by the gluon fields strength factors which form their respective Lorentz jet sum rules. In this paper, we deduce from the meson jet sum rule the remaining mesons, the J/Ψ particle with the exact mass of 3096 MeV, and the Y particles with mass 9460 MeV and 4140 MeV. For the baryons, there might be the not yet found octet with mass levels in the 5 to 8 GeV energy region, with mass level splitting also in the GeV range, far higher than those in the known octet and decuplet.

In our recent paper [1], we advanced a projection theory that provides the SU(n)xL representation, where n is 2 and 3, from the 5D homogeneous space-time manifold. This theory also provided us not just the standard model of quarks by Gell-Mann [2], but also the explicit forms of the meson-gluon fields and baryon-gluon fields. These fields together with gauge invariant constraint on the hadrons produce the mass levels observed in the octet meson representations, as well as the known octet and decuplet representations for the baryons. These mass levels are shown in the following Eight-Fold-Way diagrams below:

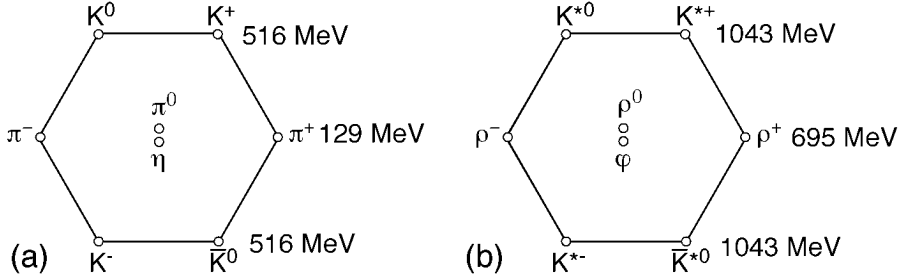

Figure 1. Meson Octets (a, b) gluon fields generated mass levels. The center meson mass not labeled.

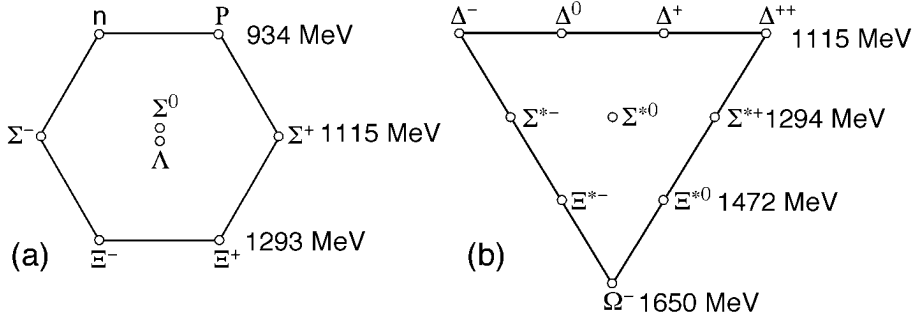

Figure 2. Baryon Octet (a) and Decuplet (b) gluon fields generated mass levels.

In paper [1] it was discussed that the fine mass splitting within each mass level was due to the bare quark masses as well as the effective Coulomb potentials (more correctly the vector potentials) between the hadron's quark constituents. These detailed fine mass splittings are however not our focal point for this article, and it suffices to point out that the theory requires that the constituent quarks of the hadron must lie on the quantum orbit as provided by the quantized flux loop, which guarantees gauge invariance. This strict restriction implies that the constituent quark states must obey Bohr-Sommerfeld quantization as well. It is this double quantum restriction that completely fixes the gluon contribution to the hadron mass. The Bohr-Sommerfeld quantization for the ground state circular model can be expressed by the following equation:

$$pr = \hbar \qquad (1)$$

where $p$ is the total rest mass relativistic momentum of the quark constituents, $r$ is the quantum orbit radius. It should be remembered that the quark mass came from the conformal projection $P_1$ of the Lorentz metric, thus it can be viewed as a phase transformation of the momentum vector. It is this feature that allows us to obtain the hadron bare quark mass contribution $m^*$ as just the simple summation of the quark constituents rest mass. Thus very much like the gauge confinement requirement of the fractional charges the hadron charge is the sum of its quark constituents. Hence, because of this Bohr-Sommerfeld quantization, we can connect the gluon contribution to the hadron mass as demonstrated for the baryons in eq. (4.20) of ref. [1], and shown pictorially in fig.1. of ref.[1]. In the gluon mass contribution to the hadron, the loop radius $r$ in the gluon potentials remains a parameter, except that it must be related to the hadron size. But because all hadrons are extremely small, its size estimation necessarily has a huge error bar. We will now show that this $r$ value can be determined via experiments. As a demonstration, we shall use the experimental data available from the study of the proton. It was found by energy conservation consideration, that the mass contribution to the proton mass due to its quark constituents (uud) can be estimated to be 11 MeV [3]. However, according to our theory, the total quark mass $m^*$ is $(5/3)m$, where $m$ is deduced as 52.5 MeV. In fact this does not mean the experimental value is much smaller, actually the experimental value is much larger than the theory predicts. As shown in ref. [1], the contribution of the quark mass to the hadron satisfies a quadratic sum rule because of the Lorentz metric, therefore it was deduced in ref.[1], that $m^*$ for the proton only provided a 4 MeV addition to the mass level value provided by the gluon potential. This difference between the experimental data and the theory is in fact the result of relativistic effects.

Before going further, we need to study how relativistic mass correction is also valid for quarks, derived from $P_1$. First the relativistic mass expression in Special Relativity comes directly from the Lorentz metric and is unchanged for either $P_0$ or $P_1$. Thus it means that the fractional quark masses satisfy the same relativistic correction for very energetic quarks. With these preliminary comments, we can now rewrite eq.(1) in terms of the $m^*$ and its velocity $v$:

$$r = (\hbar/m^*c)/\{C[C^2-1]^{0.5}\} \tag{2}$$

where C=m*(relativistic)/m*(rest mass)=$1/[1-(v/c)^2]^{0.5}$.

Eq. (2), allows us to obtain $r$ if the relativistic mass correction is known. It is precisely this feature that the comparison between the experimental determined relativistic $m^*$ contribution to the proton mass and the theories rest mass contribution, that we can determine $r$ accurately. The numerical value of $r$ that we obtained from the proton data with C=11/4 is $\hbar c/(3.64 m^*)$ or 6.18 fermi, very much in agreement with the known proton size. Which means these quark constituents are highly relativistic, with $v$=0.932c, and hence is very important to the size of the proton. This relativistic feature must be also true for all quarks in all hadrons. From this calculation it is shown that the proton mass obtained from the theory has no adjustable parameters, a clear result that no other theories can claim.

Previously [1], we forwarded a static model to show that the quark constituents in the neutron contain a residual attractive Coulomb potential, that would reduce the loop radius $r$, and hence increase the gluon potential contributed mass. It is equally consistent to forward a dynamic model, where we take the current carried by the quark constituents' charge $e^* = +e$ in the proton for contributing to a magnetic reducing flux, due to Lenz's Law, as effecting a '$r$' increase for the proton as compared to that of the neutron, and thus leading to a higher gluon generated mass of the neutron as compared to that of the proton. In fact, based on this, we can equally well obtain the gluon generated mass for the neutron just from its mass. We get M=936.7 MeV. Since the baryon-gluon potential varies as $1/r^3$, hence by comparison to the proton M and $r$ value we get straight forwardly the $r$ ratio: $r$(proton)/$r$(neutron)=$[936.7/934.5]^{1/3}$=1.001. Although the $r$ difference is only 0.1%, its contribution to the mass change is more than the differences in the quark constituents' rest mass. In fact this demonstrates the importance of the $e^*$ contribution to the individual hadron mass. Lastly, with the neutron $r$ value, we deduce the relativistic factor C from eq.(2), as 3.06, which is higher than the 2.75 for the proton. This too indicates why the total neutron quark constituents actually are more energetic than those of the proton even though their rest masses are exactly the reverse.

Because the baryon's SU(3) representations according to SU(3) should have 2 octets and 1 decuplet, but only baryons in one octet have been found and carefully measured. It is therefore our objective here to calculate and predict by using the theory of predicting the mass levels for the remaining not yet found heavy baryon octet so that an effort can be made to try to find them. In fact, this knowledge might shed light on the gamma-gamma channel of a possible energy anomaly in the range of 120 GeV detected by CERN and Fermi lab. [4]

In the 5D projection theory the strengths of the gluon fields are proportional to the product of the fractional charges square of the quark currents. Due to this proportionality, it was found that these gluon field strengths satisfy a Lorentz (jet) sum rule. [see ref.[1]] For the mesons we have

$$(1/9)^2 + 4(2/9)^2 + 4(4/9)^2 = 1 \qquad (3)$$

and for the baryons

$$(1/27)^2 + 6(2/27)^2 + 12(4/27)^2 + 8(8/27)^2 = 1 \qquad (4)$$

For the meson levels in the octets, because mesons consist of a pair of quarks, therefore only two strength factors are responsible for the mass levels within one octet. Hence it is clear that two octets must exist to account for all the three strength factors in eq. (3) which was shown in our paper [1]. It is further interesting to point out that the Eight-Fold-Way representation also implies that there must be the existence of a singlet meson state. We can deduce this singlet meson from the sum rule. As the two octets employed combined strength factors of $(1/9)^2 + 3(2/9)^2$, for the first octet, and $[(2/9)^2 + (2/9)^2] + 3[(4/9)^2 - (2/9)^2]$ for the second octet, therefore subtracting from the sum rule terms, we have a remaining strength factor for the singlet equal to $2(2/9)^2 + (4/9)^2$. Now using the corresponding mass equivalence to the factor $(1/9)^2$ corresponding to 129 MeV from the pions, a value that might be slightly large, because of the neglect of the Coulomb contribution. Irrespectively taking this value, the meson singlet should have a mass of 3,096 MeV, which is exactly the mass of the J/Ψ particle [5]. There is however another heavy meson, the Y, [6] with a mass of 9460 MeV. According to the standard model, there are no restrictions on forming a meson from (qqqq). But no such convincing (qqqq) meson evidence has been reported. It is our contention that may be the Y is still a (qq) meson but has a gluon potential generated by the (qqqq) intermediate currents. If that is the case, then this gluon potential will have a shorter range dependence of $1/r^4$, rather than those in the known octets of $1/r^2$. Thus it is not possible to deduce its coupling strength factor with knowledge of just the Y particle mass. The interesting point remains that such a heavy Y particle does not violate the theory, although this interpretation would put the Y meson outside of the 2 known standard meson octets and singlet representations. Just before submitting we learned that Fermilab had announced a new discovery of a Y(4140) [7], which they believe decays into the J/Ψ and φ. The discovery of another Y(4140) allows us to fix the (qqqq) current generated gluon potential strength, which we mentioned could be responsible for the Y(9460).

First let us consider the new Y(4140). It was found that it decays into J/Ψ and φ. We note that J/Ψ(3096) and φ(1043), therefore their sum is 4140 and satisfies mass conservation. Second J/Ψ has a gluon strength of $(4/9)^2+2(2/9)^2$, while φ gluon strength is $(4/9)^2-(2/9)^2$. The gluon generated by (qqqq) current can be approximated by (qq) (qq). If we consider that it breaks into (qq)⊔(1)+(1)⊔(qq), where (1) represents the total jet sum terms, then it is obvious that it will breakup as allowed by mass conservation. The problem that now remains is, how is the (qqqq) strength factor scaled? To analyze this, let us return to the 1976 discovered Y(9460) particle. If that is the singlet associated with the second octet in the SU(3) representations, as all (qq) strength terms were employed when J/Ψ mass was determined, then this Y(9460) must come from the total jet sum terms. We have 81x81 terms. Therefore each term corresponds to 1.418 MeV after correcting for the bare quark mass and Bohr-Sommerfeld contribution. Now the Y(4140) is approximated by 81x24+12x81=36x81. Thus using the 1.418 MeV scale we get a mass of 4135 MeV. Hence everything is consistent when the bare quark mass and Bohr-Sommerfeld contribution are added to the two Y masses.

From the agreement it is obvious that the (qqqq) gluon generated potential exists, and because of the large numbers of possible gluon potentials generated by the (qqqq) currents, there should be more yet undiscovered heavy mesons, in addition to discovered heavy mesons, such as the reported B, D, and W mesons.

In the baryon representations, a baryon being the products of 3 quarks, must be split by three of the 4 strength factors in eq.(4). However it was found in the theory that both, the known nucleon octet as well as the decuplet, only involve the first three coupling strength factors. The strength factor $(8/27)^2$ did not play a role in any of these mass levels. Because of the absence of this strength factor, it might have led to the speculation that the top quark could be super-heavy. With the introduction of the extra baryon gluon field and the three emotions (see ref.[1]), there must exist another baryon octet with mass levels all depending on this $(8/27)^2$ strength factor. Since the mass level terms were known from our calculations for the nucleon octet and the decuplet, the remaining strength terms can be obtained from subtraction from the Lorentz sum rule terms given by eq.(4). We get the following:

$$3(A)+5(B)+3(C) \qquad (5)$$

where A=$(4/27)^2$, B=$[(8/27)^2-2(2/27)^2]$ and C=$[(8/27)^2-2(1/27)^2]$, respectively.

These remaining strength terms obviously have to produce an octet and a singlet. There are many choices for the representations. We will illustrate just one such choice here. If the lighter double charge state baryons are in the first (+2 and -2) and third mass levels (-2* and +2*) and have masses proportional to A+B+C, and the heavier mass level is in the center level and is given by A+2B+C, then the singlet will be proportional to B. This heavier middle mass level should consist of +1, 0, -1 charges, with another 2e charged baryon at the center of the octet, such that the 0 charge at the center is the superposition of +2 and -2 charges. Under such an assignment this heavy octet contains a total of $7|2e|$ charges, thus in addition to the +2e state that is in the decuplet completes the 8 choices for $|2e|$ states. If the lighter mass level is proportional to A+B+C, then by using the mass number corresponding to the strength factors calculated in our paper [1], we get the two mass levels for the first assignment to be 8.443 GeV and 5.956 GeV, respectively. It should be pointed out that these baryons are far heavier than we might have guessed from known baryons. In fact the mass level splitting is also very large. Instead of a couple of hundreds of MeV, it is roughly 2.5 GeV or higher. Hence we can not expect lower energy accelerators being able to produce these baryons, and therefore this might be the reason why they have remained undiscovered. In fact, because of their very heavy level masses, the fine mass splitting due to the quark bare rest masses as well as the effective Coulomb potentials in these baryons are, relative to the mass level, extremely small, even though the constituent quark velocity might be extremely relativistic and will produce a huge relativistic mass. One can therefore even regard each mass level as 4 fold mass-degenerate. If a p-p beam is employed for its creation, then it will most likely create simultaneously all 4 degenerate baryon particles. In another word, we should expect to find them generated from the beam energies of 24 GeV, and 34 GeV, respectively. If we further consider pair production from the beam then we will be looking at 48 GeV, and 68 GeV. The sum of both simultaneous creations would require an energy in excess of roughly 120 GeV as the lighter baryons would necessarily have some kinetic energies, and no sharp resonance peak can exist. The 120 GeV anomaly observed was mentioned as a possible sign of the effect from the Higgs Boson? [4]. These particles are definitely producible in the CERN super-collider. Further, let us caution, the singlet we suggested has a mass of 2.492 GeV? It should be noted however, the strength responsible for the singlet has actually not been fixed, and its choice will be critical to nailing down these baryon octet mass levels and hence also the possible energy anomaly detectable. None the less, it is likely a mass level of A+2B+C, which is equivalent to a mass level of 8.443 GeV, would be present in this octet irrespective of its exact other mass levels.

One last remark. Although baryons in this heavy baryon octet are much heavier than the proton, it does not imply that its quark constituents are necessarily heavier. In fact all these heavier mass values come purely from the gluon strength factor. According to our projection theory, equal fraction charges must have equal fraction rest mass. Even though its relativistic mass might very well be different, as its angular momentum around the heavy baryon center is constrained by the Bohr-Sommerfeld quantization. There appears no need for the top quark rest mass to be different from the up quark. Unless we find evidence that a certain hadron mass cannot be computed from our theory, we do not feel that a conjecture for the top quark to be super heavy is warranted.